\documentclass[twocolumn,preprintnumbers,amsmath,amssymb]{revtex4}
\usepackage{graphicx}
\usepackage{dcolumn}
\usepackage{bm}

\begin{document}

\title{Spin dependent resonant tunneling through 6 micron diameter double barrier resonant tunneling diode}
\author{Z. L. Fang, P. Wu, N. Kundtz, and A. M. Chang}
\affiliation{Department of Physics, Duke University, Durham, North
Carolina 27708}
\author{X. Y. Liu and J. K. Furdyna}
\affiliation{Department of Physics, University of Notre Dame, Notre
Dame, Indiana 46556}
\date{\today}

\begin{abstract}
A vertical resonant tunneling diode (RTD) based on the paramagnetic
Zn$_{1-x-y}$Mn$_y$Cd$_x$Se system has been fabricated with a pillar
diameter down to $\sim$6 $\mu$m.  The diode exhibits high quality
resonant tunneling characteristics through the electron sub-band of
the quantum well at a temperature of 4.2 K, where a clear phonon
replica was observable in addition to the primary peak. Both peaks
show a giant Zeeman splitting in an applied magnetic field.
Employing a self-consistent real-time Green's function method, the
current-voltage characteristic was simulated, showing good agreement
with the measured result.
\end{abstract}

\maketitle

The II-VI paramagnetic semiconductor system has emerged as one of
the most promising for spintronic devices due to the presence of the
unique giant Zeeman splitting.\cite{1,2} For example, Slobodskyy and
Maximov {\it et al.} have reported voltage-controlled spin selection
in ZnMnSe magnetic RTDs and micro-patterned RTDs.\cite{3,4} In
principle, the II-VI system should allow easy tunability via gating,
once a solution for the dielectric leakage problem is found, as was
done in non-magnetic III-V systems.\cite{5} Up to now, ZnMnSe is the
typical II-Mn-VI system of choice since ZnSe is lattice matched to
GaAs substrate with only 0.27\% in-plane mismatch.\cite{3,4} On the
other hand, the partial substitution of Zn with Cd to form
Zn$_{1-x}$Cd$_x$Se alloys allows the band gap to be tuned, with an
energy gap varying from 1.75 eV for CdSe to 2.71 eV for
ZnSe.\cite{2} The Zn$_{1-x}$Cd$_x$Se system is thus more versatile
and potentially important for combining optical and electrical
devices on the same chip. However, this comes at the expense of a
much larger lattice mismatch leading to a high density of
defects.\cite{6} The room temperature cubic lattice constants for
ZnSe, GaAs and CdSe are 5.67, 5.65 and 6.05 {\AA}
respectively.\cite{2} Therefore, it is challenging to integrate
paramagnetic Zn$_{1-x-y}$Mn$_y$Cd$_x$Se into the device for
spintronics application.

\begin{figure}[b]
\includegraphics[width=62mm, clip, bb=0 0 536 600, viewport=53 279 536 600]{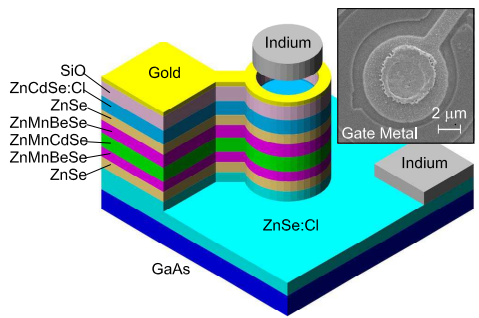}
\caption{(Color online) Schematic layer structure of the resonant
tunneling diode (gate not included) and scanning electron microscopy
image of device's top view.}\label{Fig.1}
\end{figure}

In this letter, we report the successful fabrication of a $\sim$6
$\mu$m diameter sized magnetic RTD in the Zn$_{1-x-y}$Mn$_y$Cd$_x$Se
system. Because of the small size we were able to avoid the problem
of a lack of homogeneity in the crystal often encountered in II-VI
growth. Our RTD, fabricated at a randomly chosen local region,
demonstrates excellent resonant tunneling characteristics: high
peak-to-valley current ratio in the negative differential
conductance (NDC), clear Zeeman splitting of the resonance peaks in
the tunneling current in the presence of a magnetic field, and a
distinct phonon replica. We present a complete device allowing
voltage-controlled spin selection with a self-aligned (but leaky)
gate in place. We are able to successfully explain the NDC
characteristics by employing a numerical simulation based on a
self-consistent Green's function method.

Our double barrier tunneling (DBT) crystal structure for the RTD
device was grown by molecular beam epitaxy using a GaAs (100)
substrate, as depicted in Figure \ref{Fig.1}. The DBT consists of a
600 nm {\it n}$^+$ ZnSe:Cl layer (Cl doping: 3$\times$10$^{18}$
cm$^{-3}$) for the bottom electrical contact, a 3.0 nm undoped ZnSe
spacer, a 3.0 nm Zn$_{0.62}$Mn$_{0.08}$Be$_{0.3}$Se barrier, a 6.0
nm Zn$_{0.83}$Mn$_{0.1}$Cd$_{0.07}$Se quantum well, another 5.0 nm
Zn$_{0.62}$Mn$_{0.08}$Be$_{0.3}$Se barrier and 3.0 nm undoped ZnSe
spacer, and a 65 nm {\it n}$^+$ Zn$_{0.92}$Cd$_{0.08}$Se:Cl
(2.5$\times$10$^{18}$ cm$^{-3}$) cap layer for a top contact, which
is designed to be thin, in anticipation of future devices with local
magnetic fields close to the quantum well produced by nanomagnets or
small superconductors at the surface.\cite{7} The DBT was patterned
into a $\sim$6 $\mu$m pillar by a complex, five e-beam lithography
step process. It is surrounded with a self-aligned Ti/Au gate,
deposited on a SiO$_x$ insulator as shown in the insert to Figure
\ref{Fig.1}. We note that the electrical contacts for the RTD are
made {\it ex-situ}, which is difficult for II-VI systems.\cite{8}
The contacts were produced by first evaporating a 200 nm thick
indium disk on top of the tunneling pillar, followed by annealing at
a temperature of 200 $^\circ$C for one minute. Although {\it
in-situ} deposition of the top metallic contact is often
preferable,\cite{3,4} by precisely controlling the annealing
process, we are able to consistently diffuse indium into the top
Zn$_{1-x}$Cd$_x$Se:Cl layer over a contact area of several microns
to achieve good tunneling characteristics. In addition, our device
has a 2 $\mu$m wide ridge connecting the tunneling pillar to a large
wire bonding pad. This extra connecting ridge is necessary as a
precursor to devices with local magnetic fields, where the magnetic
structures must be placed on the pillar top surface.\cite{7}

The transport characterization under magnet field was carried out by
inserting the sample into an Oxford $^4$He bath cryostat mounted
with a 6 T superconducting magnet. The magnetic field is applied
parallel to the current flow through the RTD, and positive bias
corresponds to the top, capping layer biased positively. All the
data we present were measured without gating. At present, our gate
dielectric SiO$_x$ (200 nm) is leaky, with only a low leakage window
between $-$0.8 to 0.5 V, too small to affect the tunneling
characteristics.

\begin{figure}[b]
\includegraphics[width=62mm, bb=0 0 594 841, viewport=112 350 432 570]{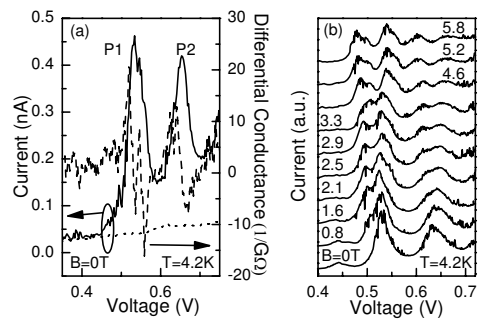}
\caption{(a) I-V (solid curve) and dI/dV-V (dashed curve) curves
with zero field at 4.2 K. Dotted curve is the I-V curve before top
contact annealing. (b) Selected I-V curves with magnetic field from
0 to 5.8 T at 4.2 K.}\label{Fig.2}
\end{figure}

The zero magnetic field current-voltage (I-V) and dI/dV-V
characteristics taken at T=4.2 K are shown in Figure \ref{Fig.2}(a).
A comparison of the I-V curves measured before and after the
precision annealing of the top indium contact indicates good contact
formation and excellent resonant tunneling behavior after annealing.
Two distinct peaks, P1 and P2, were observed in the I-V curve; P1
and P2 have peak-to-valley ratio of 3:1 and 2:1, respectively. Both
show distinct negative-differential-conductance, a key
characteristic of resonant tunneling through a DBT structure. P1 was
located at around 0.54 V and the separation between P1 and P2 is
about 105 mV. We attribute P1 to the tunneling through the second
electron sub-band in the quantum well (see following analysis). The
separation between P1 and P2 enabled us to attribute P2 to the
LO-phonon assisted in-barrier tunneling.\cite{3} The sharpness of
this replica further indicates the high quality of RTD. We ruled out
tunneling through the bonding pad as the source of the resonance
peaks. Firstly, the thin ridge turned out to be insulating because
the etched side walls contain traps that prevent the doping charge
in the top {\it n}$^+$ Zn$_{1-x}$Cd$_x$Se:Cl layer (65 nm thick)
from conducting. Secondly, cutting off the 6 $\mu$m pillar by
cutting the ridge removed the two peaks P1 and P2.

In Figure \ref{Fig.2}(b) we show the I-V characteristics under
different magnetic fields. P1 and P2 each split into two peaks due
to the giant Zeeman splitting. This splitting is repeatable and
reversible for both field directions. At the maximum field ($\sim$6
T) both have splitting magnitude $\sim$60 mV. This similar magnetic
behavior further lends support to the identification of P2 as the
LO-phonon replica of P1.

\begin{figure}[t]
\includegraphics[width=62mm, bb=0 0 594 841, viewport=134 167 456 587]{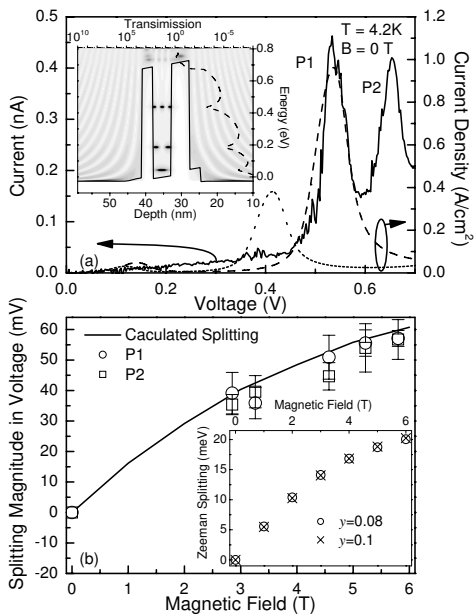}
\caption{(a) Comparison of I-V curves between experiment (solid
curve) and simulation. Dotted curve is the simulated result of a 6
nm well, dashed curve is for a 5 nm well. Insert shows the
conduction potential profile (solid curve), local density in spray
scale and transmission (dashed curve) at 0 V. (b) Comparison of the
field dependence of splitting magnitude of tunneling peaks between
experiment (circles and squares) and simulation (solid curve). The
insert is the field dependence of Zeeman splitting energy of the
conduction band in II-Mn-IV with Mn composition ({\it y}) equal to
0.08 and 0.1.}\label{Fig.3}
\end{figure}

A self-consistent real-time Green's function method was employed to
simulate the primary peak (P1) resonant tunneling behavior.\cite{9}
We use 22\% valence band offset (VBO) of ZnBeSe over ZnMnSe\cite{10}
to estimate the VBO of the barrier and well layers, which gives
around 670 meV conduction band potential in the well. The simulated
I-V curve at 4.2 K using the nominal quantum well width of 6 nm
indicates that the lowest two tunneling peaks should occur at lower
bias voltages than P1 (dotted curve in Figure \ref{Fig.3}(a)).
Tunneling peak position is mainly affected by the Fermi level
(carrier concentration) and the well width. Since the carrier
concentration in highly doped ZnSe:Cl is temperature
independent,\cite{11} the difference between simulation and
experiment is likely caused by fluctuations in the width. Indeed,
the resonant peak position was found to vary in different parts of
the crystal. With the well width as the only adjustable parameter,
we fitted to a modified value of 5 nm. All other parameters such as
electron effective mass $m^*$, dielectric constant $\varepsilon$ and
the energy gap $E_g$ were found experimentally or in the
literature.\cite{12} The simulated I-V characteristic based on the 5
nm well width has the second peak position in good agreement with
the experimental location of P1 (dashed curve in Figure
\ref{Fig.3}(a)). Therefore we attribute P1 to tunneling through
second electron sub-band. Figure \ref{Fig.3}(a) insert shows the
resultant simulated band structure, local state density and
transmission at zero source-drain voltage. The reason tunneling
through the first sub-band was not observed may be explained by the
extremely low transmission on a high current background.

The magnetic field dependence of the Zeeman splitting in the
conduction band in a II-Mn-VI semiconductor can be expressed
as:\cite{2,13} ${\Delta}E=xN_0{\alpha}s_0B_s[gs{\mu}_B/k_B(T+T_0)]$;
here, {\it x} is the Mn concentration; $N_0\alpha=0.26$ eV is the
{\it $sp$}-{\it d} exchange constant; $g$, $s$, $\mu_B$ and $k_B$
are the g-factor, manganese spin, Bohr magneton, and Boltzmann
constant, respectively. $B_s$ represents the Brillouin function of
spin $s$. The value of $s_0=1.16$ (0.95) and $T_0=2.41$ K (2.70 K)
for Mn composition of $y=0.08$ (0.1) are achieved by interpolating
values taken from the literature.\cite{14} Assuming conduction band
splitting is only significant in the barriers and well, a
two-current model to simulate each spin channel independently is
employed while neglecting the spin-coupling and scattering
disorder.\cite{15} The simulated splitting of the tunneling current
peak through the second electron level versus magnetic field is
plotted as solid line and is compared with the splitting of P1
(circles) and P2 (squares) in Figure \ref{Fig.3}(b). The simulated
result is in good agreement with experimental data.

For a RTD with negligible series resistance, the energy difference
$E$ in the well can be related to the voltage difference ${\Delta}V$
by lever arm factor  $\delta$: ${\Delta}E=e{\Delta}V/{\delta}$.
$\delta$ originates from the fact that only part of the applied
voltage is dropped on the emitter. Our RTD has a thinner emitter
barrier which further lowers the voltage drop compared to the
symmetric case, resulting in larger splitting. Comparing the
$\sim$20 meV Zeeman splitting of the electron level at 6 T field
with the $\sim$60 mV splitting observed in the resonant tunneling
peaks, we find $\delta=3$. Thus, we deduced a phonon energy of 35
meV from the P1 and P2 separation of 105 meV, close to the known LO
phonon energy of $\sim$31 meV in ZnBeSe alloy.\cite{16}

In conclusion, we report the successful fabrication of micron sized
magnetic RTD based on the paramagnetic ZnMnCdSe system with a
self-aligned gate. Clear NDC, Zeeman splitting of resonant tunneling
peaks, as well as a phonon replica were demonstrated. We demonstrate
good control on {\it ex-situ} Ohmic indium contacts, which is
usually difficult to achieve in II-VI system. Our results suggest
that further work to tune the system electrically may be possible
following improvements in gate performance.

Acknowledgement: This work was supported in part by NSF
DMR-02105191.

\end{document}